\documentstyle[epsf]{aipproc}

\begin{document}
\thispagestyle{empty}

\vspace*{-.5in}

\font\fortssbx=cmssbx10 scaled \magstep2
\hbox to \hsize{{\fortssbx University of Wisconsin - Madison}
\hfill\vtop{
\hbox{\bf MADPH-00-1198}
\hbox{October 2000}}}
\title{High Energy Neutrino Astronomy:\\ Towards Kilometer-Scale Detectors\footnote{Talk presented at the International Symposium on High Energy Gamma Ray Astronomy, Heidelberg, June 2000.}}

\author{Francis Halzen}

\address{\unskip\medskip
Department of Physics, University of Wisconsin, Madison, WI 53706}

\maketitle

\section{Introduction}
Of all high-energy particles, only neutrinos can directly convey
astronomical information from the edge of the universe---and from
deep inside the most cataclysmic high-energy processes.
Copiously produced in high-energy collisions, travelling at the velocity
of light, and not deflected by magnetic fields, neutrinos meet the basic
requirements for astronomy. Their unique advantage arises from a
fundamental property: they are affected only by the weakest of nature's
forces (but for gravity) and are therefore essentially unabsorbed as
they travel cosmological distances between their origin and us.

 Many of the outstanding mysteries of astrophysics may be hidden from
our sight at all wavelengths of the electromagnetic spectrum because of
absorption by matter and radiation between us and the source. For
example, the hot dense regions that form the central engines of stars
and galaxies are opaque to photons. In other cases, such as supernova
remnants, gamma ray bursters, and active galaxies, all of which may
involve compact objects or black holes at their cores, the precise
origin of the high-energy photons emerging from their surface regions is
uncertain.  Therefore, data obtained through a variety of observational
windows---and especially through direct observations with neutrinos---may be
of cardinal importance.

The sun is an intense source of electron neutrinos ($\nu_e$), albeit of
relatively low energy. Solar neutrino astronomy
began with first experiments in the mid-1960s; today there are
five complementary neutrino detectors viewing the nuclear reactions in
the core of the sun and, at the same time, studying the fundamental
properties of neutrinos. The sun remained the only astronomical object
studied with neutrinos until neutrinos from Supernova 1987A were
observed. These are still the only two sources marking the
astronomical neutrino spectrum.

Suggestions to use a large volume of water
for high-energy neutrino astronomy were
made as early as the 1960s~\cite{reines}.
In this case, a muon neutrino ($\nu_\mu$)
interacts with a hydrogen or oxygen nucleus in the water and produces a
muon travelling in nearly the same direction as the neutrino. The blue
\u{C}erenkov light emitted along the muon's $\sim$kilometer-long trajectory is
detected by strings of photomultiplier tubes deployed deep below the surface.
DUMAND, a pioneering project located off the coast of
Hawaii, demonstrated that muons could be detected by this
technique~\cite{DUMAND}, but
the planned detector was never realized. A detector composed of about
ninety photomultiplier tubes (or optical modules) located deep in Lake
Baikal was the first to demonstrate the detection of neutrino-induced
muons in natural water~\cite{baikal}. The European collaborations ANTARES~\cite{antares} and NESTOR~\cite{NESTOR} plan to deploy large-area detectors in the Mediterranean Sea within the next few years.  The NEMO Collaboration is conducting a site study for a future
kilometer-scale detector in the Mediterranean~\cite{NEMO}.

The AMANDA collaboration, situated at the U.S. Amundsen-Scott South Pole
Station, has strikingly demonstrated the merits of natural ice as a
\u{C}erenkov detector medium~\cite{B4,albrecht}. In 1996, AMANDA was able to observe
atmospheric neutrino candidates using only 80 eight-inch photomultiplier
tubes~\cite{B4}. With 1997 data, with 302 optical modules instrumenting approximately 6000 tons of ice, AMANDA extracted
several hundred atmospheric neutrino events.  AMANDA was the first neutrino
telescope with an effective area in excess of 10,000 square meters for TeV muons.
In January 2000, AMANDA-II was completed. It consists of 19 string with a total of 677 OMs arranged in concentric circles, with the ten strings from AMANDA forming the central core of the new detector. AMANDA has met the key challenge of neutrino astronomy: it has developed a
reliable, expandable, and affordable technology for deploying a
kilometer-scale neutrino detector named IceCube.

IceCube and the detectors to be located in the Mediterranean Sea will
be complementary in several respects. First and foremost, they will cover
different portions of the sky. (From its location at the South Pole,
IceCube observes the northern sky.) Because of the different properties
of Antarctic ice and ocean water---ice scatters light more but absorbs it
less---IceCube should have better energy resolution and a larger
effective area for neutrino detection, while the deep sea detectors should
have somewhat higher pointing resolution. Finally, one of the key
challenges in neutrino astronomy has been the ability to deploy and
maintain optical modules far below the surface; here, too, ice and
ocean are essentially different.

\section{Scientific Goals}
The many questions a high-energy neutrino telescope can address
naturally begin with astronomy, both nearby in the universe and far
away. Fascinating issues in particle physics and other branches of
science are accessible as well.

Neutrino telescopes will investigate the engines which power active galaxies, the
nature of gamma-ray bursts, and the origin of the highest-energy cosmic
rays. They will search for galactic supernovae, for the births of the
supermassive black holes which power quasars, and for the annihilation
products of halo cold dark matter particles (WIMPS and supersymmetric
particles). They can perform coincidence experiments with Earth- and
space-based gamma-ray observatories, cosmic ray detectors, and even with
future gravitational wave detectors such as LIGO.

For guidance in estimating expected signals, we make
use of the observed energy in high energy cosmic-ray protons
and nuclei as well as in known sources of non-thermal, high-energy
gamma radiation.  Some fraction of cosmic rays will interact
in their sources, whatever they may be, to produce pions.  These interactions
may be hadronic collisions
with ambient gas or photoproduction with intense photon fields
near the high-energy sources.  In either case, the neutral pions decay to
photons while charged pions include neutrinos
among their decay products with spectra related to the observed
$\gamma$-ray spectra. Estimates based on this relationship show
that a kilometer-scale detector is needed to see the neutrino
signals~\cite{snowmass}.

The observed fluxes of cosmic rays and gamma rays set the scale of neutrino telescopes. If  we, for instance, assume that gamma ray bursts are the cosmic accelerators of the highest energy cosmic rays, one can calculate from textbook particle physics how many neutrinos are produced when the particle beam coexists with the observed MeV photons in the original fireball. We thus predict the observation of order 10--100 neutrinos of PeV energy per year in a detector with a kilometer-square effective area. The rate is somewhat smaller when assuming instead that the observed cosmic ray beam originates near supermassive black holes at the center of active galaxies. Interaction of the cosmic rays with the abundant UV light in the galaxy is the source of the observed neutrinos. For detailed reviews, see reference~\cite{gaisser}.

TeV $\gamma$-rays may be produced by radiative processes from accelerated electrons, whereas neutrinos must be produced by hadronic processes.  Therefore, high-energy neutrino astronomy has the potential to discriminate
between hadronic and electromagnetic models for the intriguing TeV emission from objects such as supernova remnants, gamma-ray burst sources and active galactic nuclei. With this capability, neutrino telescopes may resolve unambiguously one of the oldest puzzles in science: the origin of cosmic rays. TeV photon fluxes determine, independently from the cosmic ray flux, the scale of the neutrino telescope. If one assumes that the observed TeV gamma rays emitted by blazars and supernova remnants are the decay products of neutral pions, one can estimate the accompanying flux of neutrinos in a largely model-independent way. One anticipates neutrino rates of order a few events per year from the Crab and the observed Markarian sources to, possibly, a few hundred from Sgr A\cite{alvarez}. The conclusion is, once more, that a kilometer-scale instrument is required to confirm or rule out the hadronic origin of the highest energy photons.

With high-energy neutrino astrophysics, we are poised to open a new
window into space and back in time to the highest-energy processes in
the universe.  The potential scientific payoff of neutrino astronomy arises from
the great penetrating power of neutrinos, which allows them to emerge
from dense inner regions of energetic sources.  It necessarily follows that the expected interaction rates are small, even with a kilometer-scale
detector.  History has shown, however, that the opening of each new
astronomical window has led to unexpected discoveries.   Thus, for example,
there could be hidden particle accelerators from which only the
neutrinos escape. Yet the science these instruments may do which we cannot anticipate, may well fuel the quest for
astronomical knowledge in the next century. Unexpected discovery has followed the inauguration of most new astronomical instruments. Large reflecting telescopes on mountaintops led to the discovery of distant galaxies and an expanding universe;
radio telescopes found the cosmic microwave background; X-ray and
gamma-ray satellites have uncovered a bestiary of awesome cosmological
objects. Even the first modest solar-neutrino observatory revealed a paradox
about the nature of fundamental interactions which is yet to be fully explained.
No one can predict all that a high-energy neutrino observatory
will find in the sky---except that it is very likely to amaze us.

\section{Large Natural \u{C}erenkov Detectors}

\begin{figure}[h]
\centering
\hspace{0in}\epsfxsize=2.2in\epsffile{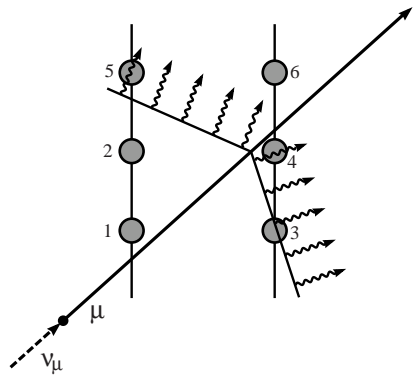}

\caption[]{The arrival times of the \u{C}erenkov photons in 6 optical sensors
determine the direction of the muon track. \label{fig:wf+oms}}
\end{figure}

The first generation of neutrino telescopes, launched by the bold decision of
the DUMAND collaboration to construct such an instrument, are
designed to reach a large telescope area and detection volume for a neutrino
threshold of order 10~GeV. This relatively low threshold permits calibration of
the novel instrument on the known flux of atmospheric neutrinos.  The
architecture is optimized for reconstructing the \u{C}erenkov light front radiated
by an up-going, neutrino-induced muon. Only up-going muons made by neutrinos
reaching us through the Earth can be successfully detected. The Earth is used
as a filter to screen the fatal background of cosmic ray muons. This makes
neutrino detection possible over the lower hemisphere of the detector. Up-going
muons must be identified in a background of down-going, cosmic ray muons which
are more than $10^5$ times more frequent for a depth of 1$\sim$2 kilometers.
The method is sketched in Fig.~\ref{fig:wf+oms}.

The optical requirements of the detector medium are severe. A large absorption
length is required because it determines the spacings of the optical sensors
and, to a significant extent, the cost of the detector. A long scattering
length is needed to preserve the geometry of the \u{C}erenkov pattern. Nature has
been kind and offered ice and water as adequate natural \u{C}erenkov media. Their
optical properties are, in fact, complementary. Water and ice have similar
attenuation length, with the role of scattering and absorption reversed. Optics seems, at present, to drive the evolution of ice and water
detectors in predictable directions: towards very large telescope area in ice
exploiting the long absorption length, and towards lower threshold and good
muon track reconstruction in water exploiting the long scattering length.

\subsection{Baikal, ANTARES, Nestor and NEMO:\break Northern Water}

Whereas the science is compelling, the real challenge is to develop a reliable,
expandable and affordable detector technology. With the termination of the
pioneering DUMAND experiment, the efforts in water are, at present, spearheaded
by the Baikal experiment\cite{baikal}. The Baikal Neutrino Telescope is
deployed in Lake Baikal, Siberia, 3.6\,km from shore at a depth of 1.1\,km. An
umbrella-like frame holds 8 strings, each instrumented with 24 pairs of 37-cm
diameter {\it QUASAR} photomultiplier tubes (PMT). Two PMTs in a pair are
switched in coincidence in order to suppress background from natural
radioactivity and bioluminescence. Operating with 144 optical modules since
April 1997, the {\it NT-200} detector has been completed in April 1998 with 192
optical modules (OM). The Baikal detector is well understood, and the first
atmospheric neutrinos have been identified.

The Baikal site is competitive with deep oceans, although the smaller absorption length of \u{C}erenkov light in lake water requires a somewhat denser spacing of the OMs. This does, however, result in a lower threshold which may be a definite
advantage, for instance for oscillation measurements and WIMP searches. They
have shown that their shallow depth of 1 kilometer does not represent a serious
drawback. By far the most significant advantage is the site with a seasonal ice
cover which allows reliable and inexpensive deployment and repair of detector
elements from a stable platform.

\begin{figure}[t]
\centering\leavevmode
\epsfxsize=3.3in\epsffile{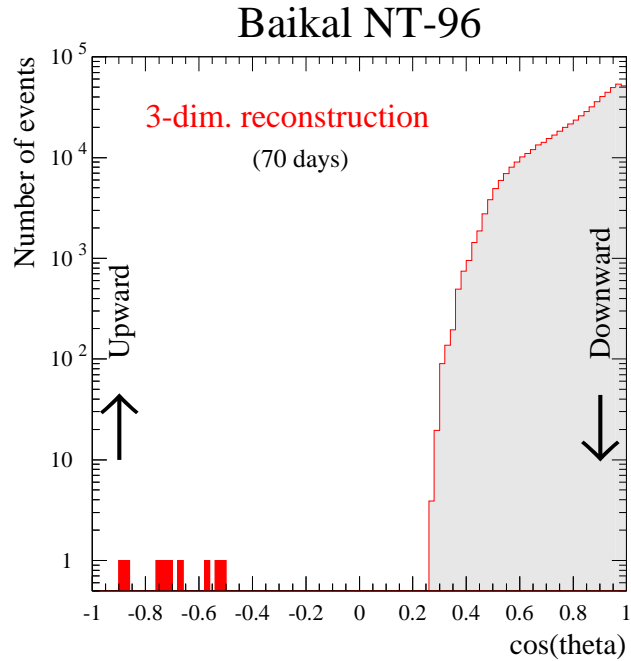}

\caption[]{Angular distribution of muon tracks in the Lake Baikal experiment
after the cuts described in the text. \label{fig:baikal}}
\end{figure}

With data taken with 96 OMs only, they have shown that atmospheric muons can be
reconstructed with sufficient accuracy to identify atmospheric neutrinos; see
Fig.~\ref{fig:baikal}. The neutrino events are isolated from the cosmic ray muon background
by imposing a restriction on the chi-square of the \u{C}erenkov fit, and by
requiring consistency between the reconstructed trajectory and the spatial
locations of the OMs reporting signals. In order to guarantee a minimum lever
arm for track fitting, they only consider events with a projection of
the most distant channels on the track larger than 35 meters. This does, of
course, result in a higher threshold.

In the following years, {\it NT-200} will be operated as a neutrino telescope
with an effective area between $10^3 \sim 5\times 10^3$\,m$^2$, depending on
energy. Presumably too small to detect neutrinos from extraterrestrial sources,
{\it NT-200} will serve as the prototype for a larger telescope. For instance,
with 2000 OMs, a threshold of  $10 \sim 20$\,GeV and an effective area of
$5\times10^4 \sim 10^5$\,m$^2$, an expanded Baikal telescope would fill the gap
between present underground detectors and planned high threshold detectors of
cubic kilometer size. Its key advantage would be low threshold.

The Baikal experiment represents a proof of concept for deep ocean projects.
These have the advantage of larger depth and optically superior water. Their
challenge is to find reliable and affordable solutions to a variety of
technological challenges for deploying a deep underwater detector. Several
groups are confronting the problem; both NESTOR and ANTARES are developing
rather different detector concepts in the Mediterranean.

The NESTOR collaboration\cite{NESTOR}, as part of a series of ongoing
technology tests, is testing the umbrella structure which will hold the OMs.
They have already deployed two aluminum ``floors", 34\,m in diameter, to a
depth of 2600\,m. Mechanical robustness was demonstrated by towing the
structure, submerged below 2000\,m, from shore to the site and back. These tests
should soon be repeated with fully instrumented floors. The actual detector
will consist of a tower of 12 six-legged floors vertically separated by 30\,m.
Each floor contains 14 OMs with four times the photocathode area of the
commercial 8~inch photomultipliers used by AMANDA and ANTARES.

The detector concept is patterned along the Baikal design. The symmetric
up/down orientation of the OMs will result in uniform angular acceptance and
the relatively close spacings in a low threshold. NESTOR does have the
advantage of a superb site off the coast of Southern Greece, possibly the best in the Mediterranean. The
detector can be deployed below 3.5\,km relatively close to shore. With the
attenuation length peaking at 55\,m near 470\,nm the site is optically superior
to that of all other deep water sites investigated for neutrino astronomy.

The ANTARES collaboration\cite{antares} is investigating the suitability of a
2400\,m-deep Mediterranean site off Toulon, France. The site is a trade-off
between acceptable optical properties of the water and easy access to ocean
technology. Their detector concept indeed requires remotely operated vehicles
for making underwater connections. First results on water quality are very
encouraging with an attenuation length of 40\,m at 467\,nm and a scattering
length exceeding 100\,m. Random noise exceeding 50\,khz per OM is eliminated by
requiring coincidences between neighboring OMs, as is done in the Lake Baikal
design. Unlike other water experiments, they will point all photomultipliers
sideways in order to avoid the effects of biofouling. The problem is
significant at the Toulon site, but only affects the upper pole region of the
OM. Relatively weak intensity and long duration bioluminescence results in an
acceptable deadtime of the detector. They have demonstrated their capability to
deploy and retrieve a string, and have reconstructed down-going muons with 8~OMs deployed on the test string.

With the study of atmospheric neutrino oscillations as a top priority, they had
planned to deploy in 2001-2003 10 strings instrumented over 400\,m with 100
OMs. After study of the underwater currents they decided that they can space
the strings by 100\,m, and possibly by 60\,m. The ANTARES detector will consist of 13 strings, each equipped with 30 storeys and 3~PMTs per storey. The large photocathode density of
the array will allow the study of atmospheric neutrino oscillations in the range $255 < L/E < 2550
\rm~km\,GeV^{-1}$ with neutrinos in the energy range $5 < E_{\nu} < 50$~GeV.
 This detector will have an area of about $3\times 10^4\rm\,m^2$ for 1~TeV muons --- similar to AMANDA-II --- and is planned to be fully deployed by the end of 2003.

A new R\&D initiative based in Catania, Sicily has been mapping Mediterranean
sites, studying mechanical structures and low power electronics. One must hope
that with a successful pioneering neutrino detector of $10^{-3}\rm\, km^3$ in
Lake Baikal, a forthcoming $10^{-2}\rm\, km^3$ detector near Toulon, the
Mediterranean efforts will converge on a $10^{-1}\rm\, km^3$ detector possibly at the
NESTOR site\cite{spiro}. For neutrino astronomy to become a viable science
several of these, or other, projects will have to succeed besides AMANDA.
Astronomy, whether in the optical or in any other wave-band, thrives on a
diversity of complementary instruments, not on ``a single best instrument".
When, for instance, the Soviet government tried out the latter method by
creating a national large mirror project, it virtually annihilated the field.

\subsection{AMANDA: Southern Ice}

\begin{figure}[h]
\centering\leavevmode
\epsfxsize=2.5in
\epsffile{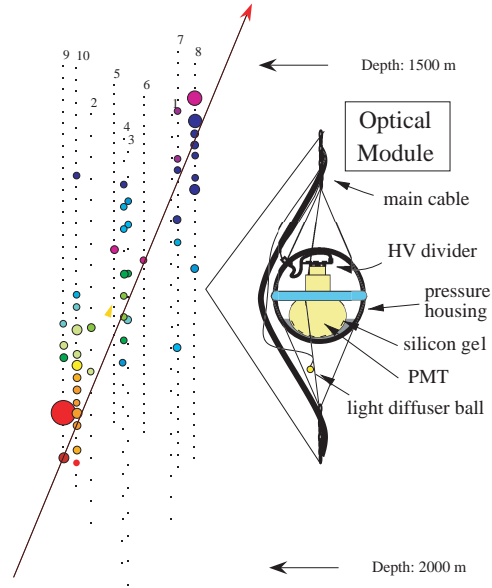}

\caption[]{The AMANDA-B10 detector and a schematic diagram of
an optical module.   Each dot represents an optical module.
The modules are separated by 20 metres in the inner strings 1-4,
and by 10 metres in the outer strings 5-10.  The colored
circles show pulses from the photomultipliers for a
particular event; the size of the circles indicate the size
of the pulse and the colors the time of the photons
arrival.  Earlier times are in red and later ones in blue.
The arrow indicates the reconstructed track of the up-going
muon. \label{fig:amanda}}
\end{figure}

Construction of the first-generation AMANDA detector was completed in the
austral summer 96--97. It consists of 302 optical modules deployed at a depth
of 1500--2000~m; see Fig.~\ref{fig:amanda}. Here the optical module consists of an 8-inch
photomultiplier tube and nothing else. It is connected to the surface by a
cable which transmits the HV as well as the anode current of a triggered
photomultiplier. The instrumented volume and the effective telescope area of
this instrument matches those of the ultimate DUMAND Octagon detector which,
unfortunately, could not be completed.

%%%%%%%%%%%% beginning of Nature text:

Depending on depth, the absorption length of blue  and UV
light in the ice varies between 85 and 225 metres.  The
effective scattering length, which
 combines the mean-free  path
$\lambda$ with the average scattering angle $\theta$ as
$\lambda   \over (1-\langle cos\theta \rangle )$,
varies from 15 to 40 metres \cite{science}.
 Because the absorption length
of light in the ice is very long and the scattering length
relatively short, many photons are delayed by scattering. In
order to reconstruct the muon track one uses maximum
likelihood methods, which take into  account the scattering
and absorption of photons as determined from calibration
measurements  \cite{B4}. A Bayesian formulation of the
likelihood, which takes into account the much larger rate of
down-going tracks relative to up-going signal, has been
particularly effective in decreasing the chance for a
down-going muon to be misreconstructed as upward-going.

Other types of events that might appear to be up-going muons
must also be considered and eliminated.  Rare cases, such as
muons which undergo catastrophic energy loss, for instance
through bremsstrahlung, or that  are coincident with other muons,
must be investigated.
To this end, a series of requirements or quality criteria,
based on the characteristic time and spatial pattern of
photons associated with a muon track and the response of the
detector, are applied to all events that, in the first
analysis, appear to be up-going muons.
 For example, an event which has a large number of optical modules
hit by photons unscattered relative to the expected Cherenkov times of the reconstructed track, has a
high quality.
  By making these requirements (or ``cuts'')
increasingly selective we eliminate correspondingly more of
the background of false up-going events while still
retaining a significant fraction of the true up-going muons,
i.e., the neutrino signal.  Two different and independent
analyses of the same set of data covering 138 days of
observation have been undertaken. These analyses yielded comparable
numbers of upgoing muons (153 in analysis A, 188 in analysis B).
Comparison of these results with their respective Monte Carlo simulations shows that they are consistent with each other in terms of the numbers of events, the number of events in common, and the
expected properties of atmospheric neutrinos.

\begin{figure}[t]
\centering\leavevmode
\epsfxsize=3.3in
\epsffile{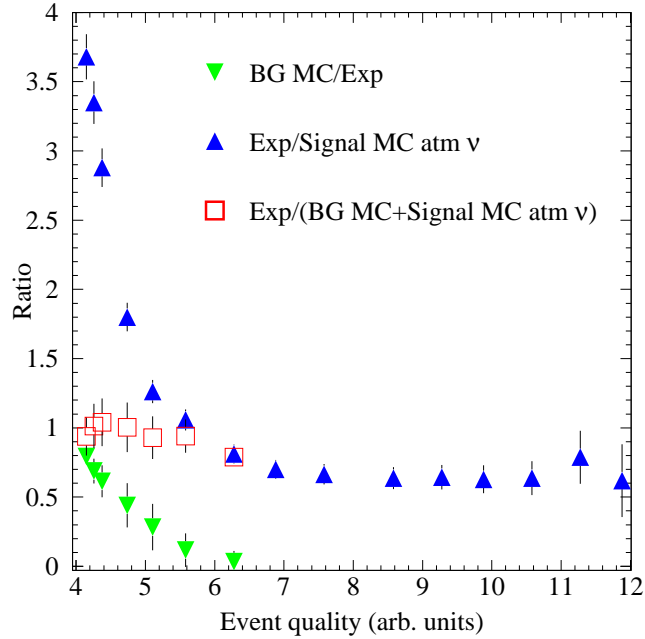}

\smallskip
\caption{Reconstructed muon events are compared to
simulations of background cosmic ray muons (BG MC) and
simulations of atmospheric neutrinos (Signal MC atm $\nu$) as a
function of ``event quality'', a variable indicating the
severity of the cuts designed to enhance the signal. Note
that the comparison is made in the form of ratios.
\label{fig:compare}}
\end{figure}

In Fig.~\ref{fig:compare}, from analysis A, the
experimental events are compared to simulations of
background and signal as a function of the (identical)
quality requirements placed on the three types of events:
experimental data, simulated up-going muons from atmospheric
neutrinos, and a simulated background of down-going
cosmic-ray muons.
 For simplicity
in presentation, the levels of the individual types of cuts
have been combined into a single parameter representing the
overall event quality, and the comparison is made in the
form of ratios.
Figure~\ref{fig:compare} shows events for which the quality level is 4 and
higher.  As the quality level is increased further, the
ratios of simulated background to experimental data and
experimental data to simulated signal both continue their
rapid decrease, the former toward zero and the latter toward
unity.  Over the same range, the ratio of experimental data
to the simulated sum of background and signal remains near
unity.  At an event quality of 6.9 there are 153 events in
the sample of experimental data and the ratio to predicted
signal is 0.7.  The conclusion is that the quality
requirements have reduced the events from misreconstructed
down-going muons in the experimental data to a negligible
fraction of the signal and that the experimental data
 behave in the same way as the simulated
atmospheric neutrino signal for events
 that pass the stringent cuts.  The remaining instrumental background is estimated at $15 \pm 7$ percent of the
signal.

The estimated uncertainty on the number of events predicted
by the signal Monte Carlo simulation, which includes
uncertainties in the high-energy atmospheric neutrino flux,
the in-situ sensitivity of the optical modules, and the
precise optical properties of the ice, is +40\% $-$50\%.  The
observed ratio of experiment to simulation (0.7) and the
expectation (1.0) therefore agree within errors.

\begin{figure}[t]
\centering\leavevmode
\epsfxsize=3.3in
\epsffile{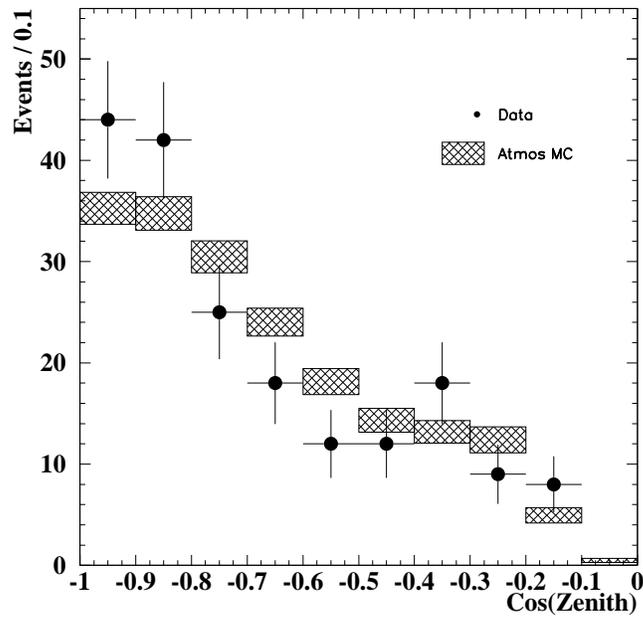}

\caption[]{Reconstructed zenith angle distribution.  The
points mark the data and the shaded boxes a simulation of
atmospheric neutrino events, the widths of the boxes
indicating the error bars. The overall normalization of the
simulation has been adjusted to fit the data.
\label{fig:zenith}}
\end{figure}

The shape of zenith angle distribution from analysis B is
compared to a simulation of the atmospheric neutrino signal
in Fig.~\ref{fig:zenith} in which the two distributions have been
normalized to each other.  The variation of the measured
rate with zenith angle is reproduced by the simulation to
within the statistical uncertainty.  Note that the tall geometry
of the detector strongly influences the dependence on zenith
angle in favor of more vertical muons.

\clearpage
Estimates of the energies of the up-going muons (based on
simulations of the number of optical modules that
participate in an event) indicate that the energies of these
muons are  in the range from 100 GeV to $\sim 1~\rm{TeV}$.  This is
consistent with their atmospheric neutrino origin.

The agreement between simulation and experiment shown in
Figures \ref{fig:compare} and \ref{fig:zenith} taken together with other comparisons of
measured and simulated events\cite{nature00}  leads to the conclusion that the up-going
muon events observed by AMANDA are produced mainly by
atmospheric neutrinos.

\begin{figure}[h]
\centering\leavevmode
\epsfxsize=4.5in
\epsffile{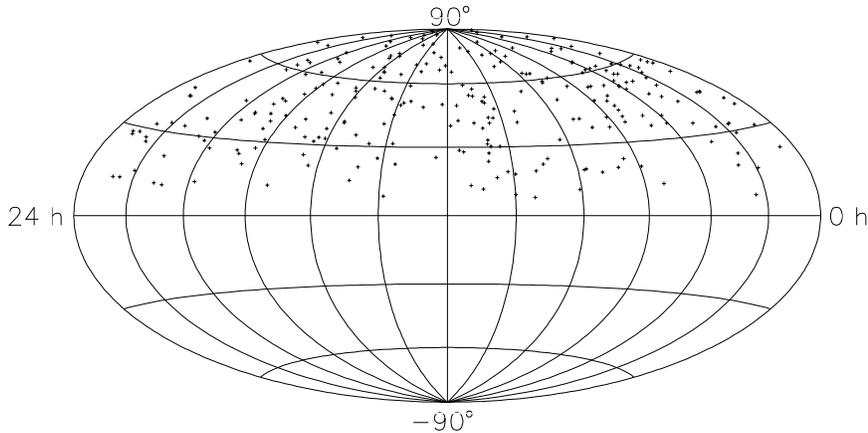}

\caption[]{Distribution in declination and right ascension of the upgoing
events on the sky. \label{fig:skyplot}}
\end{figure}

The arrival directions of the neutrinos observed in both
analyses are shown in Fig.~\ref{fig:skyplot}.
  A statistical analysis
indicates no evidence for point sources in this sample.  An
estimate of the energies of the up-going muons (based on
simulations of the number of reporting
optical modules) indicates that all events have energies consistent with an
 atmospheric neutrino origin.  This enables us to reach a level
of sensitivity to a diffuse flux of high energy extra-terrestrial
neutrinos of order $
dN/dE_{\nu} = 10^{-6} E_{\nu}^{-2} \rm\, cm^{-2}\, s^{-1}\,
sr^{-1}\,   GeV^{-1}, $ assuming an  $E^{-2}$
spectrum. At this level we would exclude a
variety of theoretical models which assume the hadronic
origin of TeV photons from active galaxies and blazars.
Searches for neutrinos from gamma-ray bursts, for magnetic
monopoles, and for a cold dark matter signal from the center
of the Earth are also in progress  and, with
only 138 days of data, yield limits comparable to or better
than those from smaller underground neutrino detectors that
have operated for a much longer period.

Data are being taken now with the  larger
array, AMANDA-II.
New surface electronics consolidates several triggering functions and adds functionality. New scalers were installed that provide millisecond resolution --- important for Supernova studies. Several technologies were deployed to evaluate their utility and readiness for future expansion to larger systems.

Yet, the fluxes of very high energy
neutrinos predicted by theoretical models or
derived from the observed flux of ultra high energy cosmic
rays are sufficiently low that a neutrino detector
having an effective area up to a square kilometer is
required for their observation and study\cite{snowmass}.
 Plans are
therefore being made for a much larger detector, IceCube,
consisting of 4800 photomultipliers to be deployed on 80
strings.  This proposed neutrino telescope would have an
effective area of $\sim 1~\rm{km^2}$,
  an energy threshold near 100 GeV
and pointing accuracy for muons  better than one degree for
high energy events.  In
conclusion, the observation of atmospheric neutrinos
reported here for AMANDA is a significant step toward
establishing the field of neutrino astronomy first
envisioned over 40 years ago.

Although neutrino telescopes are intended primarily for TeV (and higher) energy neutrino
astronomy, AMANDA and IceCube also have the potential to detect the burst of several-MeV neutrinos
from a galactic supernova and possibly even an extragalactic
burst of low-energy neutrinos associated with the birth of
a massive black hole.
By continuously monitoring the summed counting rate of its
4800 optical modules, ice detectors will monitor the sky for
such cataclysmic phenomena.  The interactions of this host of
low-energy neutrinos will be distributed uniformly
throughout the detector, a signal rising above the background noise
level.  The very low noise rate for an optical module in ice (as opposed
to the very high noise rates in ocean water) makes it possible to
continue this ``supernova watch,"
already begun by AMANDA~\cite{ralf}.

\section*{Acknowledgments}
This review has been made possible by contributions from Jaime Alvarez, Serap Tilav, and my AMANDA collaborators. This research was supported in part by the U.S.~Department of Energy under Grant No.~DE-FG02-95ER40896 and in part by the University of Wisconsin Research Committee with funds granted by the Wisconsin Alumni Research Foundation.


\begin{thebibliography}{99}
\let\sl=\it
\frenchspacing

\bibitem{reines}
K.~Greisen, Ann. Rev. Nucl. Science, {\bf 10}, 63 (1960); see also F.~Reines, Ann. Rev. Nucl. Science, {\bf 10}, 1 (1960); M.A. Markov \&
I.M. Zheleznykh, Nucl. Phys. {\bf 27}  385 (1961); M.~A. Markov in {\it
Proceedings of the 1960 Annual International Conference on High Energy Physics
at Rochester}, E.~C.~G. Sudarshan, J.~H.~Tinlot \& A.~C.~Melissinos, editors
(1960).

\bibitem{DUMAND} Cosmic Rays in the Deep Ocean, the DUMAND Collaboration (J. Babson et al.). ICR-205-89-22, Dec. 1989, 24~pp., published in Phys. Rev.~D {\bf 42}, 3613 (1990).

\bibitem{baikal}
I. A. Belolaptikov et al., Astroparticle Physics {\bf 7}, 263 (1997); V. A. Balkanov {\it et al}, Nucl. Phys. Proc. Suppl. {\bf75A}, 409 (1999).

\bibitem{antares}
E. Aslanides {\it et al}, astro-ph/9907432, 1999.

\bibitem{NESTOR} L. Trascatti, in
{\it Procs. of the 5th International Workshop on ``Topics in
Astroparticle and Underground Physics (TAUP 97)}, Gran Sasso, Italy, 1997, ed.
by A. Bottino, A.~di\,Credico, and P.~Monacelli, Nucl. Phys. {\bf B70} (Proc.
Suppl.), p.\,442 (1998).


\bibitem{NEMO} Talk given at
International Workshop on Next Generation Nucleon
Decay and Neutrino Detector (NNN99), 1999, Stony Brook, Proceedings
to by published by AIP.

\bibitem{B4}
The AMANDA collaboration, The AMANDA Neutrino Telescope: Principle of Operation and First Results, Astroparticle Physics, {\bf 13}, 1 (2000).

\bibitem{albrecht}
A. Karle, for the AMANDA collaboration, Observation of Atmospheric Neutrinos with the AMANDA Experiment, to be published in {\it Proceedings of the 17th International Workshop on Weak Interactions and Neutrinos}, Cape Town, South Africa (1999).

\bibitem{snowmass}
F. Halzen, {\it The Case for a Kilometer-Scale Neutrino Detector}, in Nuclear
and Particle Astrophysics and Cosmology, Proceedings of Snowmass\,94, R.\,Kolb and R.\,Peccei, eds.; {\it The Case for a Kilometer-Scale Neutrino Detector: 1996}, in Proc.\ of the Sixth International Symposium on Neutrino Telescopes, ed.\ by  M.\,Baldo-Ceolin, Venice (1996).

\bibitem{gaisser}
For a review, see T.K.~Gaisser, F.~Halzen and T.~Stanev, Phys. Rep. {\bf
258}(3), 173 (1995); J.G.~Learned and K.~Mannheim, to appear in Ann. Rev. Nucl. Phys; R.~Ghandi, E.~Waxmann and T.~Weiler, review talks at Neutrino 2000, Sudbury, Canada (2000).


\bibitem{alvarez} Jaime Alvarez, private communication.

\bibitem{spiro}
M. Spiro, presentation to AASC Committee of the National Academy of Sciences,
Atlanta (1999).

\bibitem{science}
P. Askjeber et al., {\it Science} {\bf 267}, 1147 (1995).

\bibitem{nature00}
E. Andres et al., submitted for publication; see also S. Barwick, Proceedings of Neutrino 2000, Sudbury, Canada (2000).

\bibitem{ralf}
The AMANDA collaboration, {\it Proceedings of the 26th International Cosmic Ray Conference}, Salt Lake City, Utah, USA (1999).



\end{thebibliography}
\end{document}